\documentclass[namedreferences]{SolarPhysics}
\usepackage[optionalrh]{spr-sola-addons} 
\usepackage{graphicx}                    
\usepackage{color}                       
\usepackage{url}                         

\newcommand{\aap}{    {\it Astron. Astrophys.}}

\newcommand{\apj}{    {\it Astrophys. J.}}

\newcommand{\pasj}{   {\it Publ. Astron. Soc. Japan}}

\newcommand{\solphys}{{\it Solar Phys.}}

\begin{document}

\begin{article}
\begin{opening}

\title{Towards Spectroscopically Detecting the Global Latitudinal 
Temperature Variation on the Solar Surface}

%
\author{Y.~\surname{Takeda}$^{1}$ and 
        S.~\surname{UeNo}$^{2}$
       }

%
\runningauthor{Y. Takeda and S. UeNo}
\runningtitle{Solar Latitudinal Surface Temperature Variation}

%
  \institute{$^{1}$ National Astronomical Observatory of Japan,
                    2-21-1 Osawa, Mitaka, Tokyo 181-8588, Japan\\
                     email: \url{takeda.yoichi@nao.ac.jp}\\ 
             $^{2}$ Kwasan and Hida Observatories, Kyoto University,
                     Kurabashira, Kamitakara, Takayama, Gifu 506-1314, Japan\\
                     email: \url{ueno@kwasan.kyoto-u.ac.jp} \\
             }

\begin{abstract}
A very slight rotation-induced latitudinal temperature variation (presumably 
on the order of several Kelvin)  on the solar surface is theoretically expected. 
While recent high-precision solar brightness observations reported its detection, 
confirmation by an alternative approach using the strengths of spectral lines 
is desirable, for which reducing the noise due to random fluctuation 
caused by atmospheric inhomogeneity is critical.
Towards this difficult task, we carried out a pilot study 
of spectroscopically investigating the relative variation of temperature ($T$) 
at a number of points in the solar circumference region near to the limb 
(where latitude dependence should show up if any exists) 
based on the equivalent widths ($W$) of 28 selected lines in the 5367--5393\AA\  
and 6075--6100\AA\ regions. Our special attention was paid 
to i) clarifying which kind of lines should be employed and 
ii) how much precision would be accomplished in practice.
We found that lines with strong $T$-sensitivity ($|\log W/\log T|$) 
should be used and that very weak lines should be avoided 
because they inevitably suffer large relative fluctuations ($\Delta W/W$). 
Our analysis revealed that a precision of $\Delta T/T \approx$~0.003 
(corresponding to $\approx$~15~K) can be achieved at best by a spectral
line with comparatively large $|\log W/\log T|$, though this can possibly 
be further improved if other more suitable lines are used. 
Accordingly, if many such favorable lines could be measured 
with sub-\% precision of $\Delta W/W$ and by averaging the resulting 
$\Delta T/T$ from each line, the random noise would eventually be reduced 
down to $\lesssim 1$~K and detection of very subtle amount of global 
$T$-gradient might be possible.
\end{abstract}

%
\keywords{Center-Limb Observations; Interior; Rotation; 
Spectral Line, Intensity and Diagnostics;
Spectrum, Visible}

\end{opening}

%

\section{Introduction}

It has occasionally been argued from theoretical considerations that latitudinal 
variation exists in the thermal structure of the solar convective envelope 
(see, {\it e.g.}, the references cited in Section~1 of Rast, Ortiz, and Meisner, 2008). 
For example, calculations by Brun and Toomre (2002) or Miesch, Brun, and Toomre (2006) 
predicted a latitude-dependent temperature variation (pole--equator temperature difference 
of $\approx 10$~K) at the tachocline to explain the solar differential rotation, 
though the extent of this variation depends upon the assumed model parameters. 
Unfortunately, not much can be said regarding the feasibility of observing 
this effect, because theoretical calculations do not extend to the surface 
layer and we have no idea about whether and how this $T$ variation
at the bottom of the convection zone is reflected in the photosphere.   
Since such a temperature difference is expected to decrease toward the surface
(Hotta, private communication; see also Hotta, Rempel, and Yokoyama, 2015), 
it might be too small to be detectable in practice.

Meanwhile, Rast, Ortiz, and Meisner (2008) carefully analyzed 
the images obtained by the {\it Precision Solar Photometric Telescope} (PSPT)
at {\it Mauna Loa Solar Observatory} and concluded a weak but systematic enhancement 
($\approx$~0.1--0.2\%) in the mean continuum intensity at the pole as compared 
to the equator, which is consistent with the theoretical prediction. 
If their conclusion is correct, a latitudinal temperature variation of 
$\approx$~2--3~K (converted by using the relation $\Delta F/F \approx 4\Delta T/T$) 
really exists in the global scale at the solar photosphere. 

Yet, it may be premature to regard this detection as being firmly established,
given its very delicate nature; {\it i.e.}, the possibility that 
the brightness enhancement may have been due to magnetic origin 
cannot be completely ruled out (as these authors noted by themselves). 
It may be noted that Kuhn {\it et al.}'s (1998) experiment using the 
{\it Michaelson Doppler Imager} (MDI) onboard the SOHO spacecraft
(Scherrer {\it et al.}, 1995)  yielded unexpectedly complex results (a temperature minimum 
between the pole and equator), which apparently contradict the theoretical
expectation as well as the PSPT observation mentioned above.
Therefore, further independent high-precision observations (preferably 
based on different techniques) would be desired.

Here, it may be worth paying attention to the spectroscopic method for diagnosing 
the surface temperature variation, which makes use of the fact that the strengths 
of spectral lines more or less change in response to temperature differences.
Actually, not a few investigators had tried to detect the global temperature 
variation in the photosphere based on this spectroscopic approach until the 1970s 
(see, {\it e.g.}, Caccin {\it et al.}, 1970; Caccin, Donati-Falchi, and Falciani, 1973; 
Noyes, Ayres, and Hall, 1973; Rutten, 1973; Falciani, Rigutti, and Roberti, 1974;
Caccin, Falciani, and Donati-Falchi, 1976;
Caccin, Falciani, and Donati Falchi, 1978; and the references therein) 
but none of them could show concrete evidence of a pole--equator temperature difference. 
Thereafter, studies in this line seem to have gone rather out of vogue 
among solar physicists and have barely been conducted so far.
Yet, some comparatively recent investigations do exist, 
though their scientific aims are somewhat different: 
Rodr\'{\i}gues Hidalgo, Collados, and V\'{a}zquez (1994) 
measured various quantities (including equivalent width) characterizing 
spectral lines on the E--W equator and N--S meridian, but no definite 
conclusion could be made regarding the existence of latitude-dependence.
Kiselman {\it et al.} (2011) searched for a latitude-dependence of line-strengths 
in the solar spectrum (to see if any aspect effect exists toward explaining 
possible differences between the Sun and solar analogs), but no meaningful
variation was detected.

Given this situation, we planned to carry out a pilot investigation, intending 
to check whether or not such an extremely subtle pole--equator temperature difference 
($\approx$~2--3~K) concluded by Rast, Ortiz, and Meisner (2008) from PSPT observations 
could ever be detected by the conventional approach using spectral lines.
Our strategy is simply to examine the behaviors of line strengths measured at 
various points on the circumference region near to the limb ({\it i.e.}, at the same radial 
distance from the disk center) such as was done by Caccin, Falciani, and Donati-Falchi (1976); 
{\it i.e.}, the latitudinal temperature gradient (if any exists) would be reflected (in theory) 
by a systematic change of line equivalent widths along the circle. 
Technically, we apply the simulated-profile-fitting method (Takeda and UeNo, 2017), 
which enables efficient and accurate measurements of equivalent widths 
no matter how many lines and points are involved.

We expect, however, that the ultimate goal is very difficult to achieve. 
Since the solar photosphere is by no means uniformly static (as approximated by the 
classical 1D model) but temporally as well as spatially variable, the fluctuation 
of line strengths inevitably caused by such inhomogeneities would be the most serious 
problem, and reducing it to as low a level as possible is essential; otherwise, it would 
easily smear out the systematic line-strength variation to be detected. In this context,
a number of lines would have to be used together to reduce the effect of random 
fluctuations by averaging. Accordingly, it is necessary in the first place 
to understand i) which kind of spectral lines are most suitable for this aim, and 
ii) to which level we can reduce the noise in practice. The purpose of this study is
to provide answers to these questions based on our experimental line-strength analysis
on the solar disk, for which we selected 28 lines of diversified properties 
(in terms of strengths and excitation conditions) as test cases.   

The remainder of this article is organized as follows. 
After describing our observations and data reduction in Section~2, 
we explain the procedures of our analysis (measurement of equivalent widths, 
conversion of line-strength fluctuation to temperature fluctuation) in Section~3.  
Section~4 is devoted to discussing the temperature sensitivity of spectral lines,
the relation between fluctuation and line strengths, and the prospect of reducing 
the noise to a required level.

\section{Observational Data}

Our spectroscopic observations were carried out on 2016 June 5, 6, 8, and 10
(JST) by using the 60~cm {\it Domeless Solar Telescope} (DST) with the
{\it Horizontal Spectrograph} at {\it Hida Observatory} of Kyoto University
(Nakai and Hattori, 1985). No appreciable active regions (sunspots, plages) 
were seen on the solar disk at this time, except for some chromospheric 
dark filaments ({\it cf.} Figure~1).  
The aspect angles of the solar rotation axis ($P$: position angle 
between the geographic North pole and the solar rotational North pole; 
$B_{0}$: heliographic latitude of the central point of the solar disk) 
were $P \approx -13^{\circ}$ and $B_{0} \approx 0^{\circ}$.
Note that we intentionally chose this period when the rotational axis was
perpendicular to the line of sight. This is because the equator appears as 
a straight line (relative to which the northern and southern hemispheres 
are symmetric) and a simple one-to-one correspondence between the coordinate   
of any disk point ($x, y$) and the heliographic latitude ($\psi$) is realized.

The target points on the solar disk are illustrated in Figure~2.
For a given position angle ($\phi$: measured anti-clockwise from the 
north rotation pole), we consecutively observed 35 points on the radial 
line (from near-limb at $r = 34/35 = 0.97$ to the disk center at $r = 0$,
where $r$ is expressed in units of the solar disk radius $R$) 
with a step of $\Delta r = 30''$~($\equiv 1/35$).
Repeating this set 24 times from $\phi = 0^{\circ}$ to 
$345^{\circ}$ with a step of $\Delta \phi = 15^{\circ}$,
we obtained spectra at 35$\times$24 points on the solar disk, 
where the slit was always aligned perpendicular to the radial direction
({\it cf.} Figure~1).
Since the disk center and the nearest-limb point correspond to 
$\cos\theta =1$ and $\cos\theta = 0.24$ $(\equiv \sqrt{1-0.97^{2}})$ 
in this arrangement ($\theta$ is the emergent angle of the ray 
measured with respect to the normal to the surface), the angle
range of $0^{\circ}\le \theta \lesssim 76^{\circ}$ is covered by our data.

In our adopted setting of the spectrograph, one-shot observation (consisting
of 30 consecutive frames with $\approx$~10--40~millisec exposure for each,
which were co-added to improve the signal-to-noise ratio) 
produced a spectrum of 0.745~\AA \,mm$^{-1}$ dispersion on the CCD detector 
covering $\approx$~25\AA\ (1600 pixels; in the dispersion direction) 
and 160$''$ (600 pixels; vertical to the dispersion direction). 
The whole observational sequence was conducted in two wavelength regions,
5367--5393\AA\ (June 10) and 6075--6100\AA\ (June 5, 6, and 8), 
which were selected because they include several representative pairs of 
spectral lines often used as temperature indicators ({\it cf.} 
Gray and Livingston, 1997a,b; Kovtyukh and Gorlova, 2000). 
 
The data reduction was done by following the standard procedures
(dark subtraction, spectrum extraction, wavelength calibration, 
and continuum normalization). The 1D spectrum was extracted by 
integrating over 188 pixels ($= 50''$; {\it i.e.}, $\pm 94$ pixels 
centered on the target point) along the spatial direction.
Given that typical granule size is on the order of $\approx$~1$''$,
our spectrum corresponds to the spatial mean of each region 
including several tens of granular cells. 
Finally, the effect of scattered light was corrected by following 
the procedure described in Section~2.3 of Takeda and UeNo (2014), 
where the adopted value of $\alpha$ (scattered-light fraction) was 
0.10 (5367--5393\AA\ region) and 0.15 (6075--6100\AA\ region) 
according to our estimation.

The S/N ratio of the resulting spectrum (directly measured from 
statistical fluctuation in the narrow range of line-free continuum) 
turned out to be sufficiently high ($\approx$~1000).
The e-folding half width of the instrumental profile (assumed to be 
Gaussian as $\propto \exp (v/v_{\rm ip})^{2}$ in this study) 
is $v_{\rm ip} \approx 1.3$~km~s$^{-1}$, which corresponds to 
a FWHM ($=2\sqrt{\ln 2} \; v_{\rm ip}$) $\approx 2.2$~km~s$^{-1}$ and 
a spectrum resolving power of $\approx 140000 (\approx c/{\rm FWHM})$ 
($c$: velocity of light). The representative disk-center spectra  
for each region (case for $\phi =0^{\circ}$) are displayed in Figure~3.

\section{Analysis}

\subsection{Measurement of Equivalent Widths}

We first selected 28 well-behaved spectral lines of diversified strengths 
(14+14 for each wavelength region) to be used for our analysis, which 
we judged as being free from any significant blending based on comparisons
with theoretically synthesized spectra. These lines (along with their basic data) 
are listed in Table~1.

Regarding the derivation of equivalent widths of these lines at various points 
on the disk, we adopted the same procedure as in Takeda and UeNo (2017) ({\it cf.} 
Section~3 therein for more details), which consists of two steps:
\begin{itemize}

\item
i) By applying the algorithm described in Takeda (1995a), the solutions of 
three parameters accomplishing the best-fit theoretical profile were 
determined: (a) $\log\epsilon$ (elemental abundance: line-strength controlling
parameter), (b) $V_{\rm los}$ (line-of-sight velocity dispersion: line-width 
controlling parameter), and  (c) $\Delta\lambda_{\rm r}$ (wavelength shift: 
line-shift controlling parameter). For the computation of theoretical profiles, 
we used Kurucz's (1993) ATLAS9 solar model atmosphere ($T_{\rm eff}$ = 5780~K, 
$\log g = 4.44$, and the solar metallicity) and $\xi$ (microturbulence)\footnote{
It has been reported that the value of microturbulence in the solar photosphere 
is rather diversified in the range of $\approx$0.5--1.4~km~s$^{-1}$ 
(see, {\it e.g.}, Section~3.2 in Takeda, 1994) and anisotropically angle-dependent 
(Holweger, Gehlsen, and Ruland, 1978). However, since this parameter plays nothing 
but an intermediary role in the present case, the resulting equivalent width does 
not depend upon its choice as long as the same value is used at step i) and step ii). 
} of 1~km~s$^{-1}$,
while the atomic data were taken from the VALD database (Ryabchikova {\it et al.}, 2015). \\

\item
ii) Then, given such established best-fit solution of $\log\epsilon$, 
we inversely computed the equivalent width ($w$) by using Kurucz's (1993) 
WIDTH9 program along with the same atmospheric model as in step i). 

\end{itemize}

The distinct merit of our method [i) + ii)] is that it does not require 
any empirical specification of the continuum level in advance ({\it cf.} Takeda, 1995a),
which enables quite an efficient derivation of line strengths in a semi-automatic manner.  
In this way, we could derive an equivalent width $w_{ij}$ at the point ($r_{i}, \phi_{j}$)
on the disk for each line, where $r_{i} = i/35$ ($i=0, 1, \cdots, 34$)
and  $\phi_{j} = 15 \times i^{\circ}$ ($i=0, 1, \cdots, 23$).\footnote{Although 
our main aim is to measure the line strengths at the circumference region
near to the limb (precisely, at $r_{32}$, $r_{33}$, $r_{34}$; {\it cf.} Figure~2),
we determined $w_{ij}$ of each line for all of the observed 840 points,
which is useful from the viewpoint of clarifying the center-to-limb variation
of $w$ as a by-product.} As examples of step i), we show in Figure~4 the 
accomplished best-fit profiles for two cases ($i=33, j=0$; near to the limb 
and $i=0, j=0$; disk center) for each line.

For the purpose of later use, we further averaged $w_{ij}$ on the same circumference 
of $r_{i}$ to derive the position-angle-averaged equivalent width ($W_{i}$)
and its standard deviation ($\sigma_{i}^{W}$) defined as
\begin{equation}
W_{i} \equiv \sum_{j=0}^{23} w_{ij}/24
\end{equation} 
and
\begin{equation}
\sigma_{i}^{W} \equiv \sqrt{\sum_{j=0}^{23}{(w_{ij}-W_{i})^2}/24}.
\end{equation}
Figure~5 depicts the center-to-limb runs of $w_{ij}$ (for 24 different $\phi_{j}$)
and $\sigma_{i}^{W}$ (plotted against $\mu_{i} \equiv \cos\theta_{i}$) 
for each of the 28 lines, while the values of $W_{0}$, $W_{33}$, and $\sigma_{33}^{W}$ 
are also presented in Table~1. 

\subsection{Fluctuations in Temperature}

We define the temperature sensitivity parameter ($k$) of the equivalent width ($w$)
of a line by the relation between the relative perturbation as $\delta w/w = k \delta T/T$, 
which is equivalent to $k \equiv d\log w/d\log T$.
Although $w$ reflects the temperature at the mean formation depth and the relevant 
value of $T$ itself is disk-position-dependent as well as line-dependent. 
we may reasonably suppose that the perturbation $\delta T/T$ is
equivalent to that of effective temperature ($\delta T_{\rm eff}/T_{\rm eff}$)
Accordingly, we numerically evaluated $k_{ij}$ corresponding to $w_{ij}$ at each point 
($r_{i}, \phi_{j}$) on the disk as follows:
\begin{equation}
k_{ij} = (w_{ij}^{+100} - w_{ij}^{-100})/w_{ij}/(200/5780),
\end{equation}
where $w_{ij}^{+100}$ and $w_{ij}^{-100}$ are the equivalent widths computed 
(with the same $\log\epsilon$ solution as used in deriving $w_{ij}$) 
by two model atmospheres with $T_{\rm eff}$ perturbed by $+100$~K 
($T_{\rm eff} = 5880$~K and $\log g = 4.44$) and  $-100$~K
($T_{\rm eff} = 5680$~K and $\log g = 4.44$), respectively.
As done for the case of equivalent width, we further averaged $k_{ij}$ over the 
position angle ($\phi_{j}$) at the same $r_{i}$ for convenience:
\begin{equation}
K_{i} \equiv \sum_{j=0}^{23} k_{ij}/24.
\end{equation} 
Again, $K_{i}$ is expressed in terms of the averaged equivalent width ($W_{i}$)
as $K_{i} \equiv \log W_{i}/\log T$. The resulting $K$ values at $r_{33}$ for 
each of the 28 lines are given in Table~1.

As the next step, we transform the relative fluctuation of local equivalent width
to that of the temperature as 
\begin{equation}
(\Delta T/T)_{ij} = (\Delta w/w)_{ij}/K_{i} = [(w_{ij} - W_{i})/W_{i}]/K_{i}.  
\end{equation}
Since the average of $(\Delta T/T)_{ij}$ over the position angle ($\phi_{j}$)
is naturally zero, its standard deviation can be calculated as
\begin{equation}
\sigma_{i}^{\Delta T/T} \equiv \sqrt{\sum_{j=0}^{23}{(\Delta T/T)_{ij}^2}/24}.
\end{equation}
The following relation holds between $\sigma_{i}^{\Delta T/T}$,
$\sigma_{i}^{W}$, and $W_{i}$ by using the absolute value of $K_{i}$:
\begin{equation}
\sigma_{i}^{\Delta T/T} =  (\sigma_{i}^{W} / W_{i})/ |K_{i}|.
\end{equation}

The runs of $(\delta T/T)_{ij}$ with the position angle ($\phi_{j}$) for three
circumferences near to the limb (at $r_{32}$, $r_{33}$, and $r_{34}$)
are illustrated in Figure~6. In Figure~7 are plotted the values of $\sigma_{33}^{W} / W_{33}$,
$K_{33}$, and $(\sigma_{33}^{W} / W_{33})/ |K_{33}| (\equiv \sigma_{33}^{\Delta T/T})$
for each line against $W_{33}$.
Further, the extent of temperature fluctuation at $r_{33}$ ($\sigma_{33}^{\Delta T/T}$)
for each line is presented in Table~1, and its line-by-line difference is graphically 
shown in Figure~8. We will discuss these results in Section~4. 

\section{Discussion} 

\subsection{Line-Dependent Temperature Sensitivity}

According to Equation~(7), the extent of temperature noise
converted from equivalent-width fluctuation of a line is determined  
by two factors: i) the parameter $K$ (indicating the $T$-sensitivity of $W$), 
and ii) the relative fluctuation of equivalent width ($\sigma^{W}/W$).

We first discuss the behavior of $K$, the values of which are considerably 
diversified (from $\approx -10$ to $\approx +10$ for those at $r_{33}$; 
{\it cf.} Figure~7b or Table~1) differing from line to line. 
This diversity is connected with the excitation property of the lower 
level from which a line originates.

The key point is whether the relevant species belongs to the major 
population or the minor population in terms of the ionization stage.
Let us consider an energy level (excitation energy $\chi_{\rm low}$)  
belonging to a species with ionization potential of $\chi^{\rm ion}$.
As explained in Section~2 of Takeda, Ohkubo, and Sadakane (2002),
the number population ($n$) of this level has a $T$-dependence of
$n \propto \exp (-\frac{\chi_{\rm low}}{kT})$ ($k$: Boltzmann constant)
if the species is in the dominant population stage, while it becomes
$n \propto T^{-3/2} \exp (+\frac{\chi^{\rm ion} - \chi_{\rm low}}{kT})
[\propto \exp (+\frac{\chi^{\rm ion} - \chi_{\rm low}}{kT})]$ (where 
we neglected the $T^{-3/2}$-dependence for simplicity because the exponential
dependence is much more significant) for the case of minor population 
species (while the parent ionization stage is the dominant population stage).
Since we may regard that the equivalent width ($W$) is roughly proportional
to the lower-level population ($n$) as far as a line is not strongly 
saturated, the parameter $K (\equiv d\log W/d\log T)$ 
is expressed as $K \approx +11604 \, \chi_{\rm low}/T \; (>0)$ (for the former
major population case) and $K \approx -11604 \, (\chi^{\rm ion} -\chi_{\rm low})/T \; (<0)$
(for the latter minor population case), where $\chi_{\rm low}$ as well as 
$\chi^{\rm ion}$ are in eV and $T$ is in K.

Regarding the 28 spectral lines (of C, Ti, V, Fe, Co, and Ni; {\it cf.} Table~1)
forming in the depth range of $-2 \lesssim \log \tau_{5000} \lesssim 0$
(see, {\it e.g.}, Figure~7b of Takeda and UeNo, 2017),
3 lines of C~{\sc i}, Ti~{\sc ii}, and Fe~{\sc ii} belong to the major population,
while the remaining lines of neutral iron-group species (Ti~{\sc i}, V~{\sc i},
Fe~{\sc i}, Co~{\sc i}, and Ni~{\sc i}) are of the minor population group,
as can be seen from Figure~9 where the population fractions of neutral, 
once-ionized, and twice-ionized stages for these 6 elements are shown
as functions of the continuum optical depth.
Then, the sign and the extent of $K$ for each line can be reasonably explained 
by the simple considerations mentioned above. This is demonstrated in Figure~10, 
where we can see that the $K$ values (at $r_{33}$) tend to be positive and proportional to 
$ +\chi_{\rm low}$  for the former major group (C~{\sc i}, Ti~{\sc ii}, 
and Fe~{\sc ii} lines), while negative and proportional to 
$ -(\chi^{\rm ion} -\chi_{\rm low})$ for the latter minor group 
(Ti~{\sc i}, V~{\sc i}, Fe~{\sc i}, Co~{\sc i}, and Ni~{\sc i} lines).

It is worth pointing out that the center-to-limb variation\footnote{ 
See, {\it e.g.}, Elste (1986), Rodr\'{\i}guez Hidalgo, Collados, and V\'{a}zquez (1994),
or Pereira, Asplund, and Kiselman (2009) for previous observational studies
on this topic.
} of the strength of a line is critically dependent upon its $K$ value.
Since the mean depth of line-formation progressively shifts to upper layers
of lower $T$ as we go toward the limb, lines with positive $K$ tend to be 
weakened while those with negative $K$ are strengthened with a decrease
in $\mu (\equiv \cos\theta)$, reflecting the
mean temperature of the layer from which photons emerge. Naturally, the degree
of this gradient becomes more exaggerated for lines with larger $|K|$.
This can be confirmed in Figure~5; for example, C~{\sc i} 5380.325
($K_{33} = +8.5$) shows a steep drop while Ti~{\sc i} 5384.630 
($K_{33} = -13.9$) exhibits a prominent rise of line strength toward the limb.\footnote{
It should be kept in mind that $K$ is not the only factor responsible for the 
center-to-limb variation in the equivalent width of a line. Actually, the change 
of the atmospheric velocity field (anisotropic microturbulence; see, {\it e.g.}, 
Appendix~1 in Takeda and UeNo, 2017) also plays a significant role for more or 
less saturated lines.}

\subsection{Relative Fluctuation of Line-Strength}

Concerning the second factor $\sigma^{W}/W$ (relative fluctuation of equivalent width),
we can recognize in Figure~7a a manifest dependence upon $W$; {\it i.e.}.
the considerably large $\sigma^{W}/W$ ($\approx$~0.05--0.1) for very weak lines 
($W \lesssim 10$~m\AA) quickly drops with an increase of $W$ and settles
at $\approx$~0.01--0.02 for medium-strength lines of $W \gtrsim 30$~m\AA.
Actually, the large value of $\sigma^{W}/W$ for the case of weak lines 
can be recognized in Figure~5 ({\it e.g.}, Fe~{\sc i} 5378.236, Co~{\sc i} 5381.770, 
Ti~{\sc i} 5384.630, Fe~{\sc i} 5385.575, Fe~{\sc i} 5392.015, Ti~{\sc i} 6092.792, 
and Ti~{\sc i} 6098.658). This means that the standard deviation ($\sigma^{W}$) defined 
by Equation~(2) is almost independent of the line strength ($W$),\footnote{
In this context, it is worth noting that error of equivalent width can be expressed
in terms of the signal-to-noise ratio, half-width, and profile-sampling step,
without any relevance to the line-strength itself ({\it cf.} Cayrel, 1988).} 
which yields a prominent enhancement of $\sigma^{W}/W$ as $W$ decreases down to 
a very weak-line level.
Accordingly, weak lines ($W \lesssim 30$~m\AA), which were preferably used in some 
previous work intending to detect subtle temperature differences ({\it e.g.}, Caccin, 
Falciani, and Donati-Falchi, 1976), should be avoided, because they inevitably 
suffer a large relative fluctuation of $W$. On the other hand, strong saturated lines
($W \gtrsim$~100--200m\AA) showing appreciable wings are also not suitable, because 
of the increased difficulty in accurately measuring the equivalent widths.   
We consider that medium-strength lines (30m\AA~$\lesssim W \lesssim$~100m\AA) 
are most favorable for this purpose.

\subsection{Toward Precisely Determining Temperature Variation}

Based on Equation~(7) along with the arguments in Section~4.1 and Section~4.2,
what should be done to suppress the noise ($\sigma^{\Delta T/T}$) for precise
temperature determination is manifest: i) to use lines of as large 
$T$-sensitivity  ({\it i.e.}, large $|K|$) as possible, and ii) to reduce
the relative fluctuation of equivalent width ($\Delta W/W$) as much as possible
by using medium-strength lines while avoiding very weak lines.

In the present pilot study, the smallest $\sigma^{\Delta T/T}$ we have achieved
at best is $\approx$~0.003 (\#11, \#16, \#17, \#20; {\it cf.} Table~1 or Figure~8),
which corresponds to $\approx 15$~K in temperature. That the dispersions of $\Delta T/T$
for these lines are notably small can be visually confirmed in Figure~6.
Although these lines are rather weak ($W \approx$~20--40~m\AA; {\it cf.} Figure~7c)
which may not necessarily be desirable from the viewpoint of strength fluctuation,
this is simply due to the fact that large $|K|$ lines happen to be rather weak 
among our 28 spectral lines (Figure~7b).
This means that we may further improve the precision 
if other more suitable lines (sufficiently large $K \approx 10$ 
and medium-strength lines expected to have small $\Delta W/W$) could be used. 

As a trial, we checked the 311 blend-free lines, which were used by Takeda (1995b) 
for studying the line profiles of the solar flux spectrum, in order to see how many
such suitable lines are found therein. Most of these lines belong to neutral 
iron-group elements ({\it i.e.},  minor population species), though a small fraction 
of them are from major population species (Fe~{\sc ii}, Ti~{\sc ii}).
The flux equivalent widths (taken from Table~1 and Table~2 of Takeda, 1995b)
of these lines are plotted against the relevant critical energy potential ({\it i.e.}, the energy 
difference between the lower level and the ground level of the dominant species) 
in Figure~11, where the upper panel shows the $W_{\lambda}^{\rm flux}$ {\it vs.} $\chi_{\rm low}$ 
relation for major population species and the lower panel is for the
$W_{\lambda}^{\rm flux}$ {\it vs.} $\chi^{\rm ion} - \chi_{\rm low}$ relation of 
minor population species. Recalling that those lines with larger values in the abscissa 
have stronger $T$-sensitivity or larger $|K|$ ({\it cf.} Figure~10),
we can see from Figure~11 that a few tens of suitable lines
(with $\chi_{\rm low} \gtrsim$~5~eV or $\chi^{\rm ion} - \chi_{\rm low} \gtrsim$~5~eV
while the strength is in the modest range of 
30~m\AA~$\lesssim W_{\lambda}^{\rm flux} \lesssim$~100~m\AA),
which presumably have $|K|$ values around $\approx 10$, can be actually found among 
these 311 lines. Accordingly, we may hope that a larger number of such lines 
would be further available if we turn our eyes to wider wavelength regions. 

Regarding the relative fluctuation of equivalent widths ($\sigma^{W}/W$), the minimum 
level we could accomplish at best in this study was $\approx$~0.01--0.02 (Figure~7a).
Presumably, there is still room for further reducing this noise.
First of all, our spectrum at each point on the disk was obtained by spatially 
averaging $50''$ along the slit direction (corresponding to several tens of granule cells) 
as mentioned in Section~2. If we could extend the spatial area (in the two-dimensional 
sense) including a much larger number of granules 
({\it e.g.}, $\approx$~$10^{3}$--$10^{4}$) to be averaged, we may expect 
a significant reduction of $\sigma^{W}/W$.
Alternatively, conducting long-lasting observations (instead of snap-shot
observations adopted in this investigation) at the same target point 
over a sufficiently large time span ({\it e.g.}, several tens of minutes, 
such as done by Kiselman {\it et al.}, 2011) would also be useful to smear 
out the temporal fluctuation especially due to solar oscillations.
Admittedly, such special observations (involving many points on the disk,
long time span, and wide wavelength coverage) may not be easy in practice. 
Nonetheless, if realized, we speculate that the noise in $W$ measured 
on the averaged data would be reduced down to the sub-\% level. 

If we could keep $\sigma^{W}/W$ well below $<0.01$ for a suitable $T$-sensitive 
spectral line with $K \approx 10$, Equation~(7) suggests that the noise in 
$\Delta T/T$ for this line would be $< 0.001$ ({\it i.e.} $< 5$~K).
Then, making use of as many ({\it e.g.} several tens to a hundred) such lines 
as possible combined, we may be able to accomplish a precision of $\lesssim 1$~K,
and the detection of a global $T$-difference of a few K would eventually be possible. 

\subsection{Other Effects Influencing Line Strengths}

Since the main subject of this study is to detect the latitudinal temperature 
variation based on the equivalent widths of spectral lines, we have focused so far 
only on the $T$-dependence of $W$, while evaluating its sensitivity by adopting a 
static plane-parallel solar atmospheric model along with the assumption of LTE. 
However, the actual solar photosphere is not so simple as to be represented by such a 
classical model and line strengths are generally affected by other factors, as has 
been investigated by previous studies  ({\it e.g.}, Jevremovi\'{c} {\it et al.}, 1993; 
Sheminova, 1993, 1998; Cabrera Solana, Bellot Rubio, and del Toro Iniesta, 2005).
We briefly discuss below whether these non-temperature effects can cause any 
confusion problem in our context.

\subsubsection{Magnetic Field}

In addition to active regions of apparently strong magnetism (spots, plages), 
magnetic fields of various forms are known to exist on the solar surface
even in quiet regions or polar regions (see, {\it e.g.}, 
Stenflo, and Lindegren, 1977; Trujillo Bueno, Shchukina, and Asensio Ramos, 2004; 
Nordlund, Stein, and Asplund, 2009; S\'{a}nchez Almeida and 
Mart\'{\i}nez Gonz\'{a}lez, 2011; Petrie, 2015). 
The effect of these photospheric magnetic fields on the strengths of spectral 
lines was studied by several investigators, especially from the viewpoint of 
impact on abundance determination ({\it e.g.}, Fabbian {\it et al.}, 2010, 2012; 
Criscuoli {\it et al.}, 2013; Moore {\it et al.}, 2015). 

The main influence of magnetic field on the equivalent width is due to
the Zeeman splitting of the line opacity profile, which generally acts 
in the direction of strengthening a line. Besides, there is an indirect effect
caused by a lowering of gas pressure under the existence of magnetic pressure,
if the field is  sufficiently strong, though many important lines of minor 
population species (such as Fe~{\sc i} lines) are not sensitive to gas pressure 
({\it i.e.}, to $\log g$) as shown by Takeda, Ohkubo, and Sadakane (2002).
We roughly checked how the Zeeman splitting corresponding to a magnetic field 
of 100~G affects the equivalent widths of 28 spectral lines (Table~1)
following the procedure described in Takeda (1993; {\it cf.} Case (c) therein), 
where our estimation gives only the upper limit of magnetic intensification 
because the polarization effect was neglected.
The resulting relative changes of local equivalent widths ($w_{33}$) at $r_{33}$ 
are plotted against $g_{\rm L}^{\rm eff}$ (effective Land\'{e} factor; {\it cf.} 
7th column in Table~1) are shown in Figure~12a, where we can see that $\delta w_{33}/w_{33}$ 
for 100~G is on the order of $\approx 10^{-3}$ and tends to increase with $g_{\rm L}^{\rm eff}$.

Generally, if localized or patchy magnetic regions are involved, 
we may presumably be able to discriminate their effects, because they would appear 
something like spurious humps. Meanwhile, the ubiquitous magnetic field in 
quiet regions would not really be a problem as far as it is position-independent, 
since what we want to detect is the latitudinal variation of line strengths. 
However, the global magnetic field caused by the rotation-induced dynamo mechanism
(see, {\it e.g.}, Ulrich and Boyden, 2005) would bring about some interference, 
because they may produce systematic large-scale variation of line-strengths,
even though the field strength is fairly weak ($\lesssim 10$~G).  
At any rate, it is desirable to monitor the existence and nature of magnetic field 
at the points where the intended observations are to be made.   

\subsubsection{Velocity Field}

The random motions of various scales in the solar photosphere affect the shape as well as 
the strength of spectral lines. In the traditional modeling of line formation  (as 
adopted in this study), this turbulent velocity field is very roughly represented
by two parameters; {\it i.e.}, ``micro''-turbulence of microscopic scale and 
``macro''-turbulence of macroscopic scale (see, {\it e.g.}, Gray, 2005).
In the framework of this dichotomous modeling, the total line strength (equivalent 
width) is affected only by the former (microturbulence), and that strong saturated 
lines are sensitive to its change while weak unsaturated lines are not.
In Figure~12b is shown how the local equivalent widths ($w_{33}$) of 28 lines are affected
by a change of the microturbulence by 0.1~km~s$^{-1}$ ($1.0 \rightarrow 1.1$~km~s$^{-1}$).
We can recognize from this figure that, while $\delta w_{33}/w_{33}$ is very small
in the regime of weak lines ($w_{33} \lesssim 10$~m\AA), it progressively grows
(up to several per cent) with an increase of $w_{33}$, as expected.

While these turbulence parameters (for both micro- and macro-turbulence)
are known to be anisotropic in the sense that they depend on the view angle ($\theta$) 
relative to the normal to the surface ({\it cf.} Holweger, Gehlsen, and Ruland, 1978; 
Takeda and Ueno, 2017), we tend to naively consider that no dependence exists upon 
the position angle ($\phi$) for a given circumference of same $\theta$; 
then, the anisotropy of this kind in the turbulent velocity field should not have 
any influence on the latitudinal $T$-dependence under question.   

However, there are some reports that cast doubts on this simple picture of
circular symmetry. For example, Caccin, Falciani, and Donati-Falchi (1976)
suggested based on their analysis of weak line profiles a very small dependence of 
photospheric turbulence velocity upon the heliographic latitude.  
Moreover, Muller, Hanslmeier, and Utz (2017) very recently found a latitude variation 
of granulation properties (size and contrast) based on high-resolution space 
observations by the {\it Hinode} satellite at the period of activity minimum (see also 
Rodr\'{\i}guez Hidalgo, Collados, and V\'{a}zquez, 1992 for a similar conclusion).
Yet, it is not clear how this latitudinal variation of granular velocity field
affects the line strengths (equivalent widths) of our concern. 
In any event, this point would have to be kept in mind. 
 
\subsubsection{Non-LTE Effect}

We assumed LTE in evaluating the $T$-sensitivity of $W$ throughout this study.
This is nothing but a simplified  approximation and the more realistic treatment 
without this assumption (non-LTE calculation) more or less changes the strengths 
of spectral lines. However, LTE is essentially a valid treatment in the context of
the primary purpose of this study, because the non-LTE effect itself
does not directly come into the problem. That is, what matters here is the 
relative variation ($\delta W/W$) in response to a change of temperature ($\delta T$), 
and $\delta W^{\rm NLTE}/W^{\rm NLTE} \approx \delta W^{\rm LTE}/W^{\rm LTE}$
practically holds even when an appreciable non-LTE effect is expected.
For example, although strong C~{\sc i} 10683--10691 triplet lines in the near infrared 
suffer a significant non-LTE effect as shown by Takeda and UeNo (2014), the values of $k$ 
($\equiv d\log W/d\log T$) computed for these lines are almost the same for both 
the LTE and non-LTE case ({\it i.e.} the differences are only a few \%). 

\section{Concluding Remarks}

Given that a very slight rotation-induced latitudinal temperature gradient 
on the order of several K is suspected on the solar surface, we carried out 
a pilot investigation to examine whether it is possible to verify the existence 
of such a global temperature change based on the strengths of spectral lines.
Since detecting a very small signal of global $T$-variation while reducing the noise 
due to random fluctuation caused by atmospheric inhomogeneity to a sufficiently 
low level must be critical for accomplishing this aim, it is important
in the first place to clarify which spectral lines are suitably $T$-sensitive
for this purpose. 

According to our trial analysis using 28 lines in 5367--5393\AA\ and 6075--6100\AA\ 
regions based on the spectroscopic observations at a number of points on the solar disk, 
we found that a specific group of lines (lines of minor population species
with large $\chi^{\rm ion} - \chi_{\rm low}$ or lines of major population species
with large $\chi_{\rm low}$) have a strong sensitivity to a change of $T$
($|\log W/\log T|$ as large as $\lesssim 10$), and the noise level can be suppressed
to $\sigma^{W}/W \approx$~0.01--0.02 by employing such a suitable line.
Since this is a precision achieved by a single line, we can further improve 
it by using a number of lines all together.

Therefore, we may hope for observations specifically designed in terms of 
the following points: i) wide wavelength coverage (availability of 
as many suitable lines as possible), ii) covering a larger spatial area 
(to include a much larger number of granules), and iii) covering a longer time span 
(to erase the temporal fluctuation especially due to non-radial oscillations).
We then could eventually accomplish a precision of $\lesssim 1$~K, and the 
detection of a global $T$-difference by a few K would be put into practice. 

In addition, we should not forget the possibility that factors other than
temperature may affect the strengths of spectral lines, such as the influence
of magnetic field (Section~4.4.1) or a latitude-dependent granular velocity field
(Section~4.4.2). For this reason, it is desirable to monitor the detailed 
conditions of observed points ({\it e.g.} the nature of magnetic regions or 
distributions of granular motions/brightness) by coordinated high-resolution 
spectro-polarimetric observations; for example, with the {\it Solar Optical Telescope} 
aboard the {\it Hinode} mission (Tsuneta et al., 2008). 
With the help of such supplementary observations, we would be able to 
eliminate confusion effects irrelevant to the global temperature variation. 

Consequently, regarding the question ``is it possible to spectroscopically 
detect the very small latitudinal temperature gradient of only $\approx$~2--3~K?'' 
our tentative answer is ``we do not think it impossible, and it may be 
accomplished by careful and well arranged observations,'' 
although this would be a considerably tough task to practice.

\begin{acks}
This work has made use of the VALD database, operated at Uppsala 
University, the Institute of Astronomy RAS in Moscow, and 
the University of Vienna.
\end{acks}
\newline
\newline
{\bf Disclosure of Potential Conflicts of Interest}\\
The authors declare that they have no conflicts of interest.

\setcounter{table}{0}
\begin{table}[h]
\tiny
\caption{Basic data and the derived results for the chosen 28 spectral lines.}
\begin{center}
\begin{tabular}
{ccrcccrcrrcrc}\hline 
\# & species & $\chi^{\rm ion}$  & class & $\lambda$ & Lower--Upper & $g_{\rm L}^{\rm eff}$ & $\chi_{\rm low}$ &
$W_{0}$ & $W_{33}$ & $\sigma^{W}_{33}$ & $K_{33}$ &  $\sigma_{33}^{\Delta T/T}$ \\
(1) & (2) & (3) & (4) & (5) & (6) & (7) & (8) & (9) & (10) & (11) & (12) & (13) \\
\hline
 1&  Fe~{\sc i}&   7.87&  m & 5373.708&  $^{3}{\rm G}_{3}^{\circ}- ^{3}{\rm F}_{4}$   &  2.00  & 4.473&  71.08&  78.81&  1.11& $ -2.75$& 0.00513 \\
 2&  Fe~{\sc i}&   7.87&  m & 5376.830&  $^{1}{\rm D}_{2}-^{3}{\rm P}_{1}^{\circ}$    &  0.75  & 4.295&  12.52&  15.88&  0.49& $ -5.19$& 0.00596 \\
 3&  Fe~{\sc i}&   7.87&  m & 5378.236&  $^{5}{\rm F}_{4}^{\circ}-??_{5}$             &$\cdots$& 5.033&   2.57&   3.37&  0.14& $ -4.93$& 0.00866 \\
 4&  Fe~{\sc i}&   7.87&  m & 5379.573&  $^{1}{\rm G}_{4}-^{1}{\rm H}_{5}^{\circ}$    &  1.00  & 3.695&  59.33&  67.52&  0.74& $ -2.62$& 0.00418 \\
 5&  C~{\sc i} &  11.26&  M & 5380.325&  $^{1}{\rm P}_{1}^{\circ}-^{1}{\rm P}_{1}$    &  1.00  & 7.685&  24.51&  13.47&  0.43& $ +8.54$& 0.00377 \\
 6&  Ti~{\sc ii}& 13.58&  M & 5381.022&  $^{2}{\rm D}_{3/2}-^{2}{\rm F}_{5/2}^{\circ}$&  0.90  & 1.566&  53.70&  59.50&  0.41& $ +0.37$& 0.01854 \\
 7&  Co~{\sc i}&   7.86&  m & 5381.770&  $^{4}{\rm F}_{3/2}^{\circ}-^{4}{\rm G}_{5/2}$&  0.70  & 4.240&   4.75&   6.42&  0.25& $ -5.50$& 0.00712 \\
 8&  Fe~{\sc i}&   7.87&  m & 5382.263&  $^{5}{\rm G}_{5}^{\circ}-^{5}{\rm H}_{6}$    &  1.08  & 5.669&  22.02&  25.17&  0.55& $ -2.40$& 0.00913 \\
 9&  Ti~{\sc i}&   6.82&  m & 5384.630&  $^{5}{\rm F}_{3}-^{3}{\rm F}_{3}^{\circ}$    &  1.17  & 0.826&   0.91&   1.56&  0.13& $-13.89$& 0.00585 \\
10&  Fe~{\sc i}&   7.87&  m & 5385.575&  $^{1}{\rm G}_{4}-^{5}{\rm H}_{4}^{\circ}$    &  0.95  & 3.695&   4.38&   6.26&  0.21& $ -7.43$& 0.00445 \\
11&  Fe~{\sc i}&   7.87&  m & 5386.333&  $^{5}{\rm D}_{3}^{\circ}-^{5}{\rm P}_{2}$    &  1.17  & 4.154&  31.42&  38.66&  0.51& $ -3.73$& 0.00355 \\
12&  Fe~{\sc i}&   7.87&  m & 5389.478&  $^{5}{\rm G}_{3}^{\circ}-^{5}{\rm G}_{3}$    &  0.92  & 4.415& 102.20& 117.26&  1.41& $ -3.05$& 0.00394 \\
13&  Fe~{\sc i}&   7.87&  m & 5392.015&  $^{3}{\rm D}_{2}^{\circ}-^{3}{\rm F}_{3}$    &  1.00  & 4.795&   5.26&   7.17&  0.33& $ -5.17$& 0.00880 \\
14&  Ni~{\sc i}&   7.64&  m & 5392.331&  $^{3}{\rm D}_{3}^{\circ}-^{3}{\rm F}_{2}$    &  2.00  & 4.154&  11.05&  14.37&  0.62& $ -4.05$& 0.01066 \\
15&  Fe~{\sc i}&   7.87&  m & 6078.491&  $^{3}{\rm D}_{2}^{\circ}-^{3}{\rm F}_{3}$    &  1.00  & 4.795&  89.33&  93.33&  1.67& $ -2.37$& 0.00753 \\
16&  V~{\sc i} &   6.74&  m & 6081.442&  $^{4}{\rm D}_{3/2}-^{4}{\rm P}_{3/2}^{\circ}$&  1.47  & 1.051&  12.31&  16.59&  0.58& $-10.57$& 0.00332 \\
17&  Fe~{\sc i}&   7.87&  m & 6082.710&  $^{5}{\rm P}_{1}-^{3}{\rm P}_{1}^{\circ}$    &  2.00  & 2.223&  31.83&  40.16&  0.70& $ -5.30$& 0.00329 \\
18&  Fe~{\sc ii}& 16.18&  M & 6084.102&  $^{4}{\rm G}_{9/2}-^{6}{\rm F}_{7/2}^{\circ}$&  0.78  & 3.200&  21.05&  21.22&  0.57& $ +3.92$& 0.00682 \\
19&  Ni~{\sc i}&   7.64&  m & 6086.282&  $^{3}{\rm D}_{1}^{\circ}-^{3}{\rm F}_{2}$    &  0.75  & 4.266&  49.41&  52.97&  0.67& $ -2.09$& 0.00609 \\
20&  V~{\sc i} &   6.74&  m & 6090.208&  $^{4}{\rm D}_{7/2}-^{4}{\rm P}_{5/2}^{\circ}$&  1.21  & 1.081&  31.11&  38.40&  0.68& $ -6.68$& 0.00267 \\
21&  Ti~{\sc i}&   6.82&  m & 6091.171&  $^{1}{\rm G}_{4}-^{1}{\rm H}_{5}^{\circ}$    &  1.00  & 2.267&  13.02&  16.74&  0.56& $ -8.15$& 0.00412 \\
22&  Ti~{\sc i}&   6.82&  m & 6092.792&  $^{3}{\rm G}_{5}-^{3}{\rm G}_{5}^{\circ}$    &  1.20  & 1.887&   3.56&   5.00&  0.49& $-10.96$& 0.00895 \\
23&  Co~{\sc i}&   7.86&  m & 6093.141&  $^{4}{\rm P}_{3/2}-^{4}{\rm D}_{3/2}^{\circ}$&  1.47  & 1.740&   8.15&  12.34&  0.58& $ -9.67$& 0.00490 \\
24&  Fe~{\sc i}&   7.87&  m & 6093.642&  $^{3}{\rm F}_{3}^{\circ}-^{5}{\rm P}_{2}$    &  0.33  & 4.608&  32.29&  36.36&  0.72& $ -3.31$& 0.00596 \\
25&  Fe~{\sc i}&   7.87&  m & 6094.372&  $^{3}{\rm F}_{2}^{\circ}-^{5}{\rm P}_{1}$    & $-0.25$& 4.652&  18.31&  20.97&  0.71& $ -4.30$& 0.00790 \\
26&  Fe~{\sc i}&   7.87&  m & 6096.664&  $^{3}{\rm F}_{2}^{\circ}-^{3}{\rm F}_{3}$    &  1.50  & 3.984&  39.34&  46.08&  0.99& $ -3.60$& 0.00598 \\
27&  Fe~{\sc i}&   7.87&  m & 6098.243&  $^{5}{\rm P}_{3}^{\circ}-^{5}{\rm P}_{3}$    &  1.67  & 4.559&  15.47&  18.75&  0.77& $ -4.58$& 0.00892 \\
28&  Ti~{\sc i}&   6.82&  m & 6098.658&  $^{1}{\rm G}_{4}^{\circ}-^{1}{\rm F}_{3}$    &  1.00  & 3.062&   4.92&   6.05&  0.49& $ -8.17$& 0.00982 \\
\hline
\end{tabular}
\end{center}
\tiny
(1) Line number. (2) Species. (3) Ionization potential (in eV). (4) ``M'': major 
population species, ``m'': minor population species.
(5) Wavelength (in \AA). (6) Designations of lower and upper levels.
(7) Effective Land\'{e} factor.
(8) Excitation potential of the lower level (in eV).
(9) Position-angle-averaged equivalent width at $r_{0}$ (in m\AA).
(10) Position-angle-averaged equivalent width at $r_{33}$  (in m\AA).
(11) Standard deviation of $W_{33}$ (in m\AA).
(12) Position-angle-averaged temperature sensitivity of equivalent width 
at $r_{33}$. (13) Standard deviation of $\Delta T/T$ at $r_{33}$.
\end{table}

\begin{figure} 
\centerline{\includegraphics[width=0.8\textwidth]{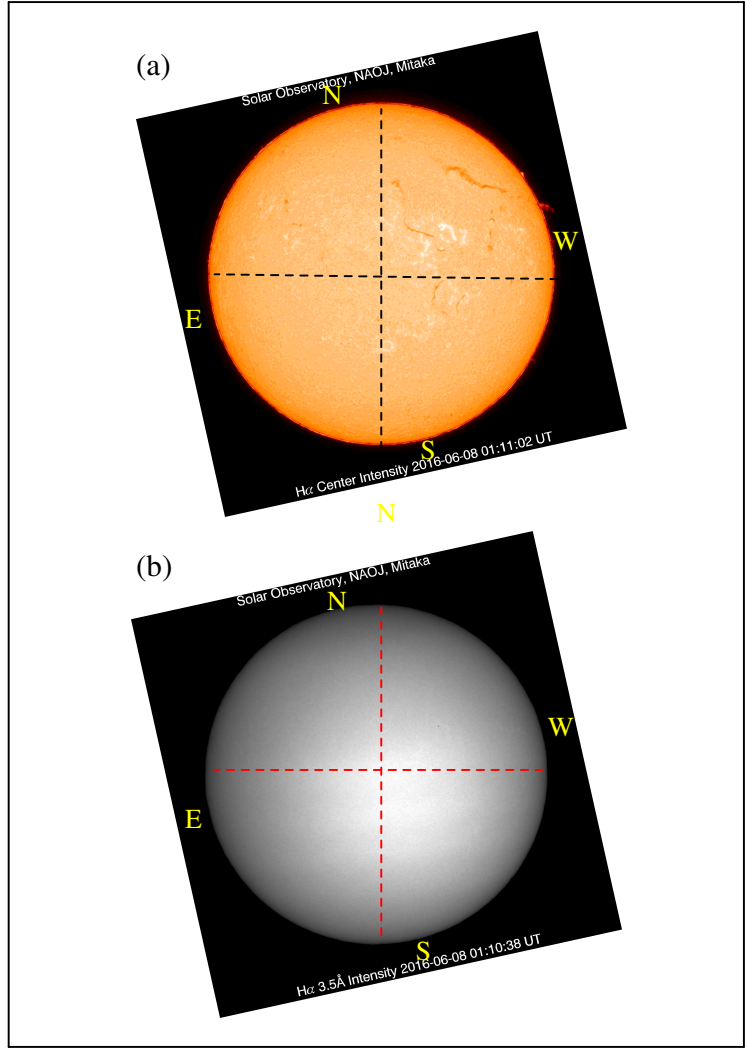}}
\caption{In the upper and lower panels are shown (a) the H$\alpha +0.0$~\AA 
(line-center) image and (b) the H$\alpha + 3.5$~\AA (nearly equivalent 
to continuum) image of the Sun on 2016 June 8, observed by 
the {\it Solar Flare Telescope} at the Solar Observatory, National Astronomical
Observatory of Japan.
The solar meridian and equator (depicted in dashed lines) 
are aligned with the vertical and horizontal directions, respectively,
in order to maintain consistency with Figure~2.
}
\end{figure}

\begin{figure} 
\centerline{\includegraphics[width=0.8\textwidth]{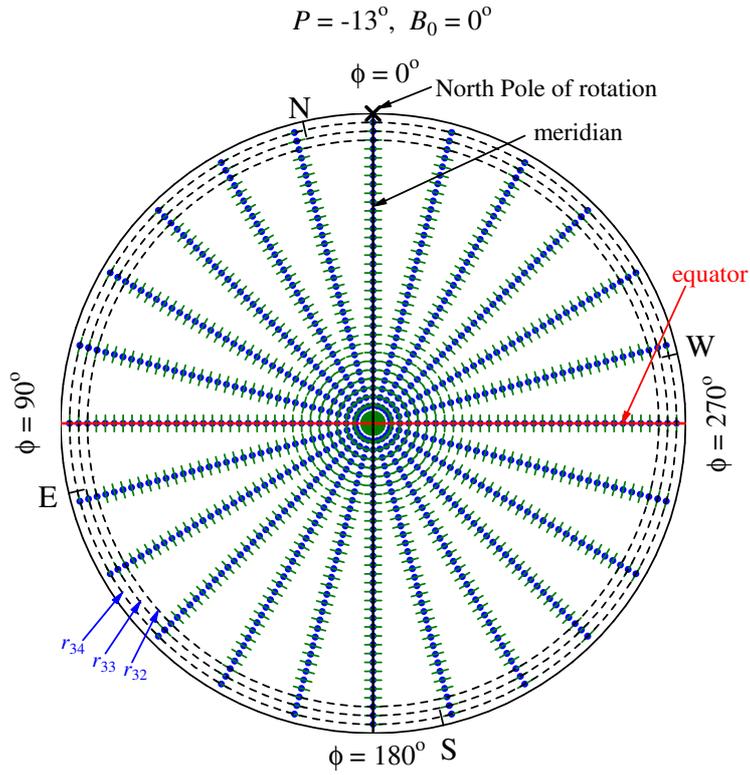}}
\caption{
Graphical description of the observed 840 (= 35$\times$24) points 
on the solar disk, at which the spectral data were taken. 
Their locations are specified by 35 radial distances from $r = 0$  
(disk center) to $r = 0.97$ (near to the limb) with a step of 
$\Delta r = 1/35 (\simeq 30'')$ in units of the solar disk radius,
and 24 position angles ($\phi$) with an increment of 15$^{\circ}$. 
The short segments (in line with alignment of the slit being 
perpendicular to the radial direction) show the integrated range ($50''$) 
in the spatial direction for each of the spectra. 
}
\end{figure}

\begin{figure} 
\centerline{\includegraphics[width=0.8\textwidth]{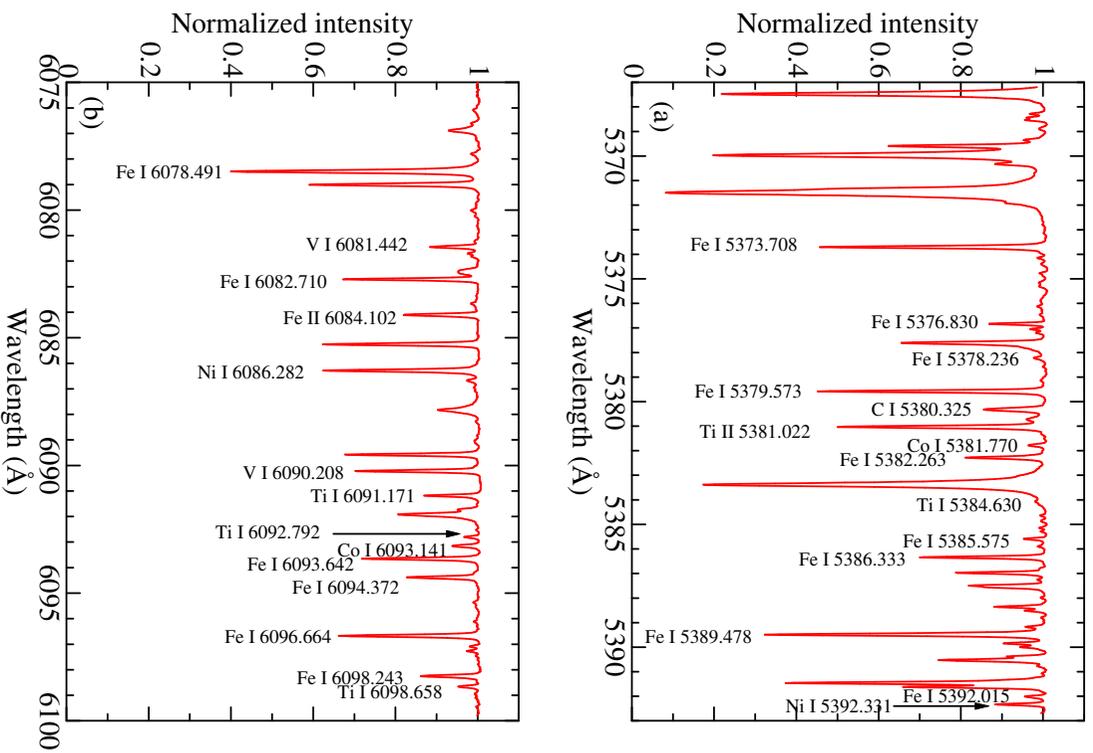}}
\caption{Examples of our spectra at the disk center ($r = 0$, $\phi = 0^{\circ}$)
in (a) the 5367--5393\AA\ region, and (b) the 6075--6100\AA\ region.
The 28 spectral lines we selected are indicated in the figure.  
}
\end{figure}

\begin{figure} 
\centerline{\includegraphics[width=1.0\textwidth]{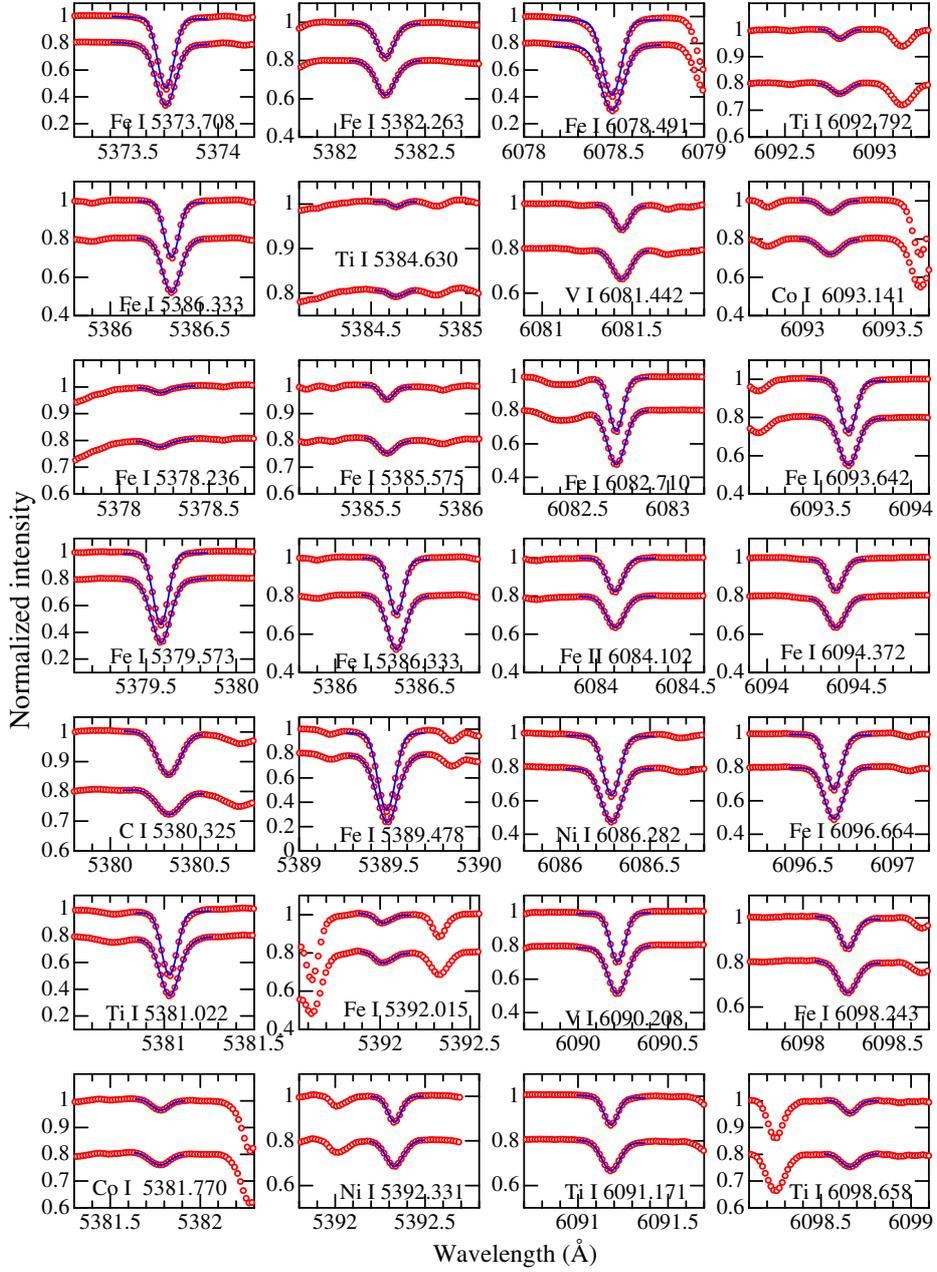}}
\caption{Comparison of the observed spectrum (open circles) and 
the fitted theoretical spectrum (solid line) for each of the 28 lines. 
In each panel, two spectra at the disk-center ($r_{0}, \phi_{0}$) 
and the near-limb ($r_{33}, \phi_{0}$) are depicted, where the latter 
is shifted downward with an offset of 0.2. 
}
\end{figure}

\begin{figure} 
\centerline{\includegraphics[width=1.0\textwidth]{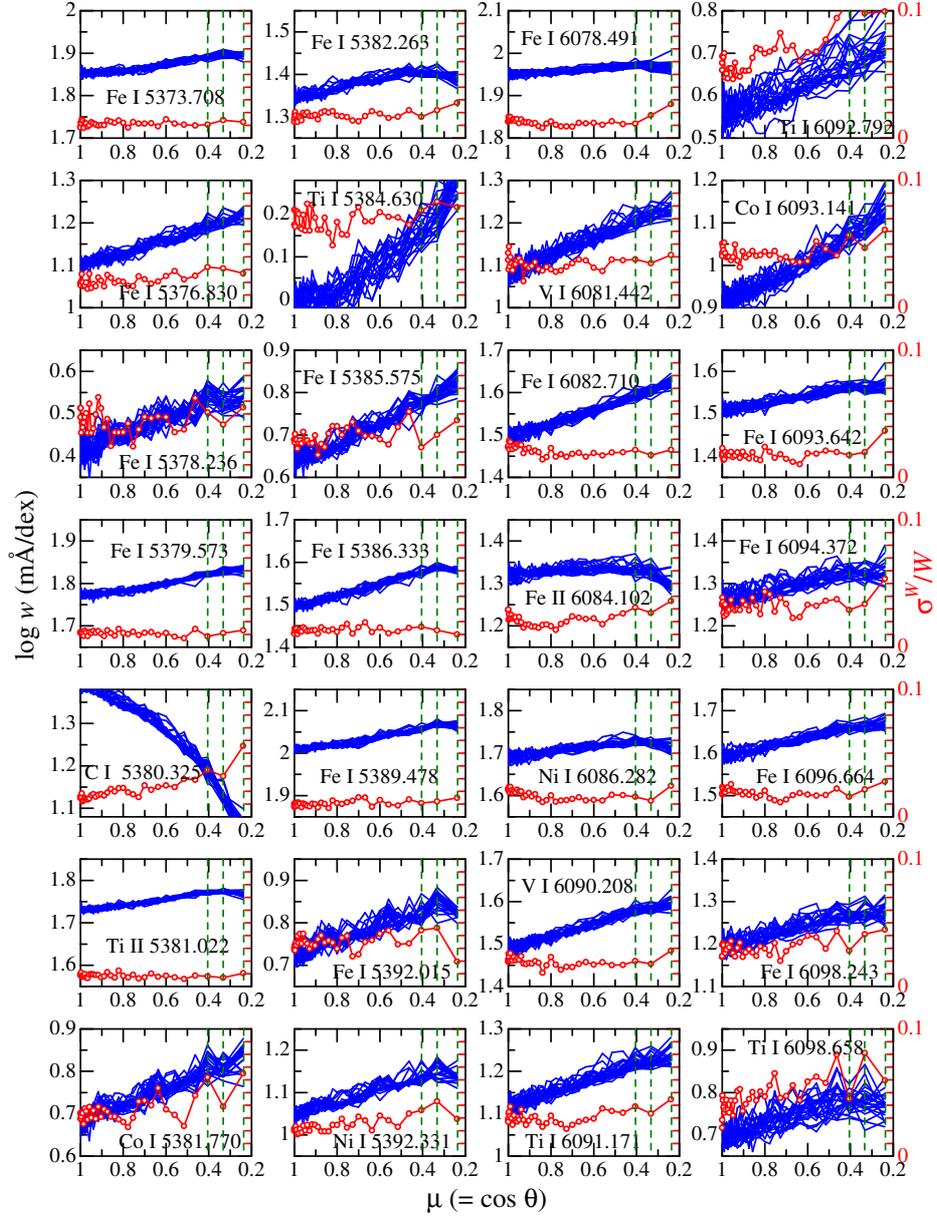}}
\caption{
Solid lines (scale on the left-hand axis): center-to-limb run of $w_{ij}$ 
[local equivalent width at ($r_{i}$, $\phi_{j}$)] with $\mu_{i}$  
($i = 0, 1, \cdots, 34$) for each of the 24 position angles 
($\phi_{j}$, $j = 0, 1, \cdots, 23$)
where $\mu_{i} \equiv \cos\theta_{i} \equiv \sqrt{1- (i/35)^2}$.
Open-circle-connected lines (scale on the right-hand axis of the rightmost panels): 
center-to-limb run of $\sigma_{i}^{W}/W_{i}$ with $\mu_{i}$, where $W_{i}$ and $\sigma_{i}^{W}$ 
are the mean $w_{ij}$ averaged over $\phi_{j}$ and its standard deviation, respectively. 
The positions of $r_{32}$, $r_{33}$, and $r_{34}$ (near-limb regions
of our concern) are indicated by the vertical dashed lines.
}
\end{figure}

\begin{figure} 
\centerline{\includegraphics[width=1.0\textwidth]{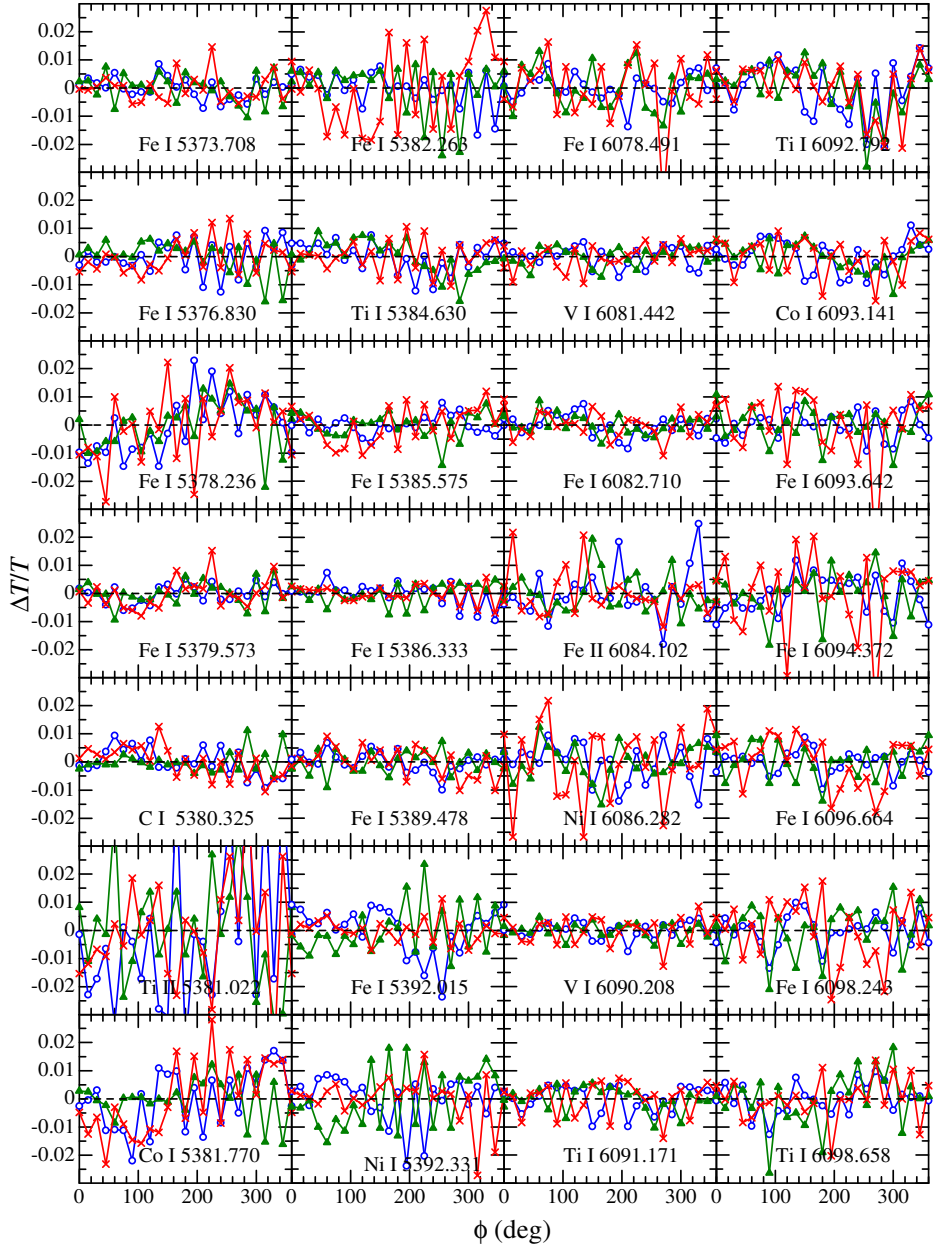}}
\caption{
Run of the temperature fluctuation ($\Delta T/T$) with the position
angle $\phi_{j}$ ($j = 0, \cdots, 23$) near to the limb at $r_{32}$ (open 
circles), $r_{33}$ (filled triangles), and $r_{34}$ (crosses), which was 
formally derived from the equivalent-width fluctuation $(\Delta w/w)_{ij}$ 
by using the $T$-sensitivity parameter ($K_{i}$).
}
\end{figure}

\begin{figure} 
\centerline{\includegraphics[width=0.5\textwidth]{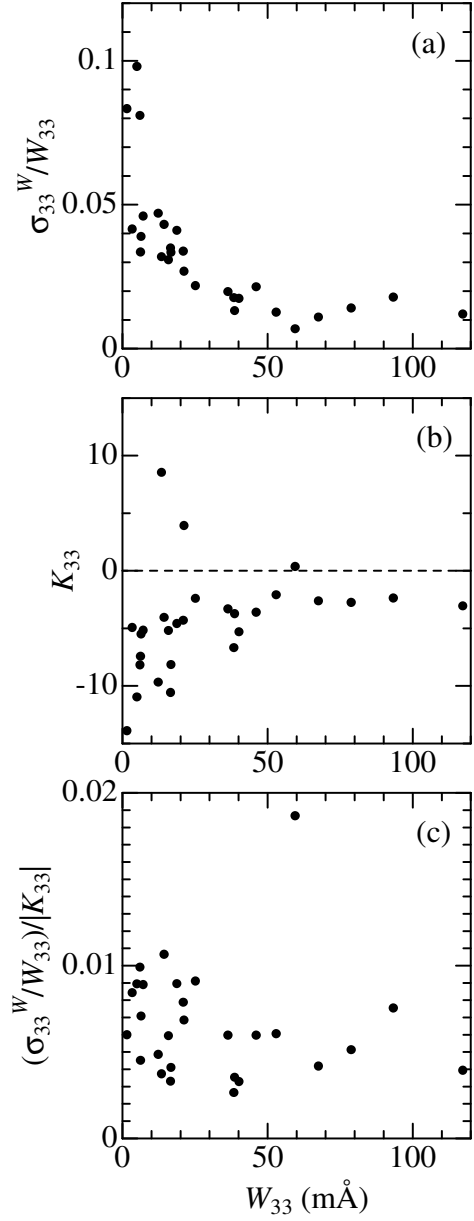}}
\caption{
Values of (a) $\sigma^{W}/W$, (b) $K$, and (c) $(\sigma^{W}/W)/|K|$ 
for each line plotted against $W$ (all these values correspond to  
$r_{33}$ near to the limb).
}
\end{figure}

\begin{figure} 
\centerline{\includegraphics[width=0.8\textwidth]{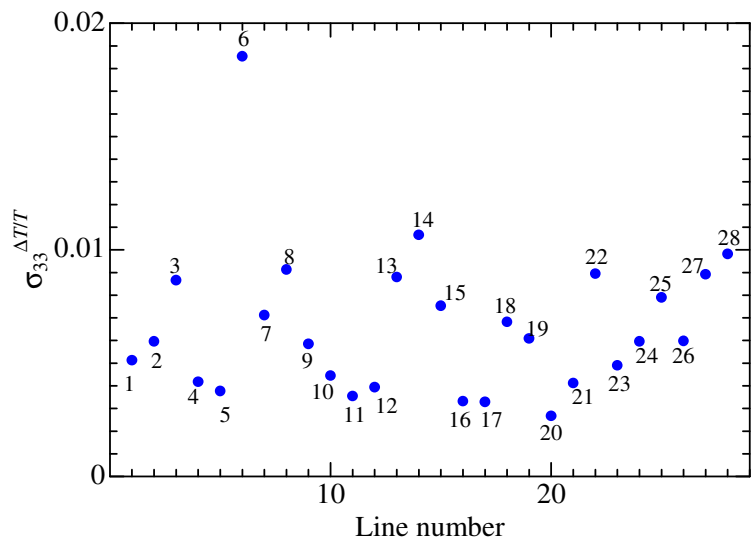}}
\caption{
Standard deviation of $\Delta T/T$ for each line (calculated over 
the circumference near to the limb at $r_{33}$) plotted against 
the line number. 
}
\end{figure}

\begin{figure} 
\centerline{\includegraphics[width=0.8\textwidth]{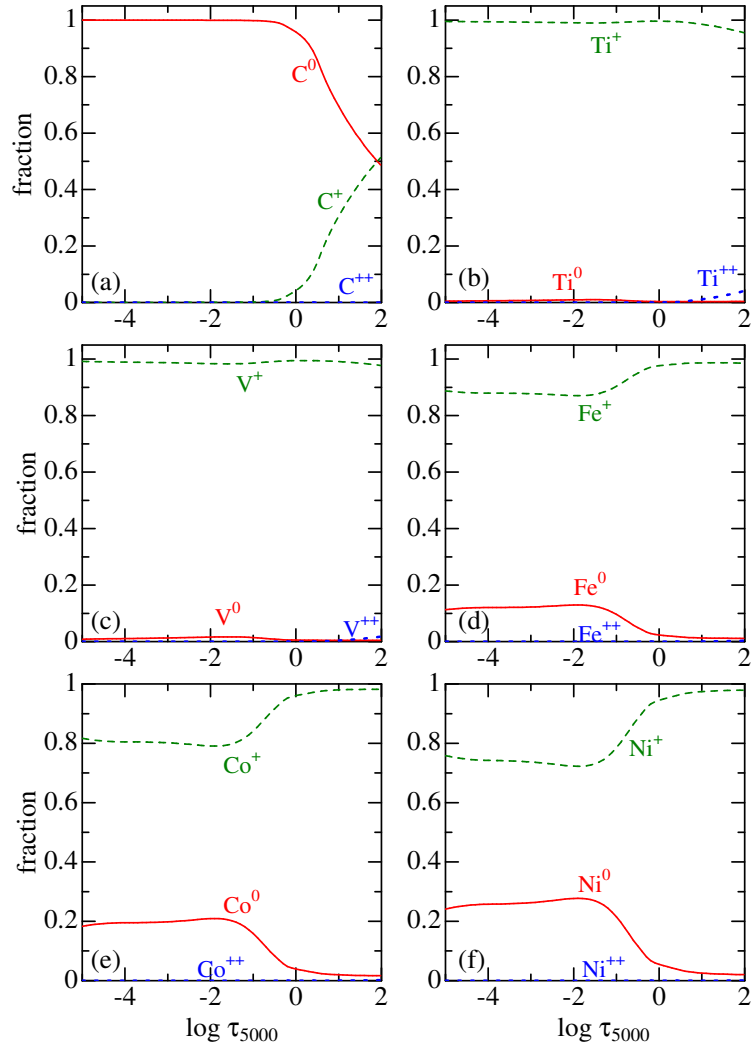}}
\caption{Number population fractions at the neutral (X$^{0}$;
solid line), once-ionized (X$^{+}$; dashed line), and
twice-ionized (X$^{++}$; dotted line) stages plotted against 
the standard continuum optical depth at 5000~\AA,  
which were computed from Kurucz's (1993) solar model atmosphere.
The panels show: (a) C, (b) Ti, (c) V, (d) Fe, (e) Co, and (f) Ni.
}
\end{figure}

\begin{figure} 
\centerline{\includegraphics[width=0.7\textwidth]{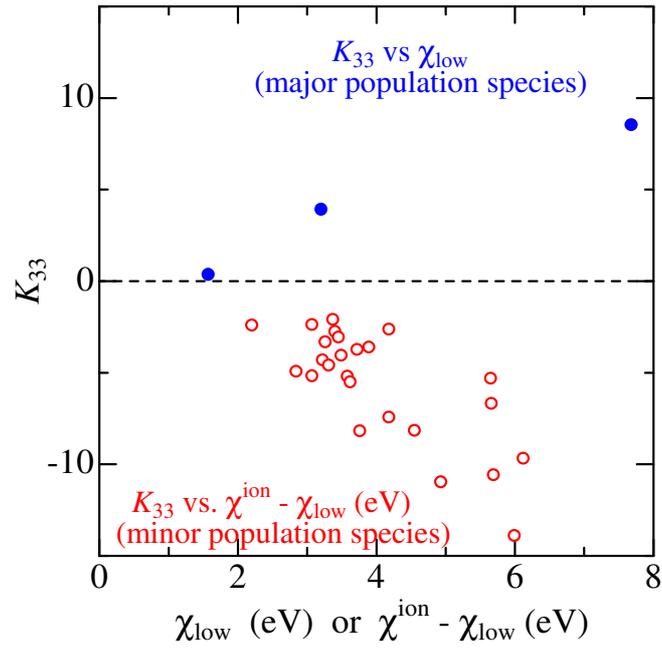}}
\caption{
The $K$ values (temperature sensitivity of the line strength)  
computed at $r_{33}$ for each line plotted against the critical potential energy. 
Filled circles: case for the major population species (C~{\sc i}, Ti~{\sc ii}, 
and Fe~{\sc ii}) with positive $K$, where the abscissa is $\chi_{\rm low}$.
Open circles: case for the minor population species (Ti~{\sc i}, V~{\sc i}, 
Fe~{\sc i}, Co~{\sc i}, and Ni~{\sc i}) with negative $K$, where the abscissa 
is $\chi^{\rm ion} - \chi_{\rm low}$.
}
\end{figure}

\begin{figure} 
\centerline{\includegraphics[width=0.7\textwidth]{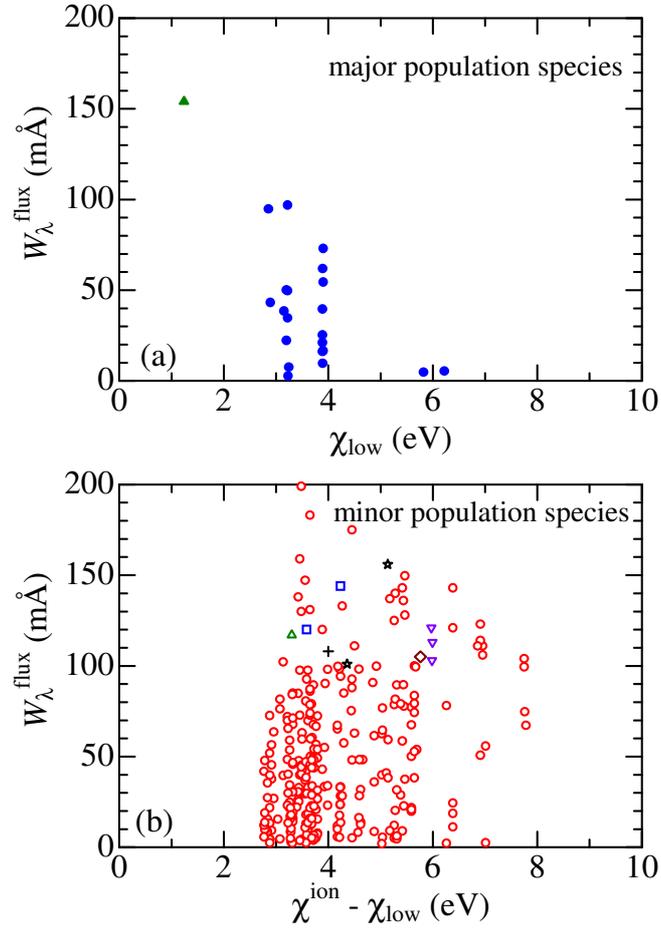}}
\caption{
The solar flux equivalent widths of the 311 lines used by Takeda (1995b)
plotted against the critical potential energy.
(a) Case of major population species where the abscissa is $\chi_{\rm low}$.
Filled triangle: Ti~{\sc ii}, filled circles: Fe~{\sc ii}.
(b) Case of minor population species, where the abscissa is 
$\chi^{\rm ion} - \chi_{\rm low}$. Open triangle: Mg~{\sc i},
open squares: Ca~{\sc i}, open downward triangles: Ti~{\sc  i},
open diamond: Cr~{\sc i}, open stars: Mn~{\sc i}, 
open circles: Fe~{\sc i}, and the cross: Ni~{\sc i}.
}
\end{figure}

\begin{figure} 
\centerline{\includegraphics[width=0.6\textwidth]{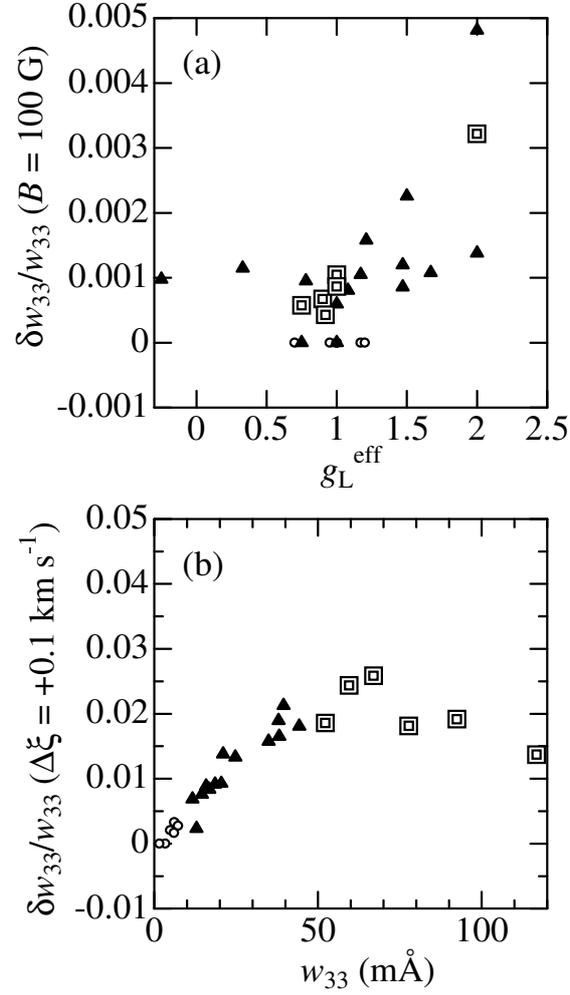}}
\caption{
(a) Computed relative variations of equivalent widths near to the limb 
($\delta w_{33}/w_{33}$; at $i=33$, $j=0$) caused by the Zeeman splitting 
in the presence of a magnetic field  of 100~G, plotted against 
$g_{\rm L}^{\rm eff}$ (effective Land\'{e} factor). 
(b) Computed relative variations of $\delta w_{33}/w_{33}$
caused by by increasing the microturbulence by 0.1~km~s$^{-1}$ 
(from 1.0 km~s$^{-1}$ to 1.1~km~s$^{-1}$), plotted against $w_{33}$.
Different symbols (and sizes) are used according to the strengths of $w_{33}$.
Small open circles: $w_{33}<$~10~m\AA,  
filled triangles: 10~m\AA $\le w_{33}<$~50~m\AA, and
large double squares: 50~m\AA $\le w_{33}$.
}
\end{figure}

\end{article} 
\end{document}